# Imaging Cyclotron Orbits of Electrons in Graphene

*Sagar Bhandari[†], Gil-Ho Lee[†], Anna Klales[†], Kenji Watanabe[¶], Takashi Taniguchi[¶],*

*Eric Heller[†], Philip Kim[†], and Robert M. Westervelt[†,\*]*

[†]School of Engineering and Applied Sciences and Department of Physics

Harvard University, Cambridge, MA 02138 USA

[¶]National Institute for Materials Science

1-1 Namiki, Tsukuba, 305-0044, Japan

**ABSTRACT:** Electrons in graphene can travel for several microns without scattering at low temperatures, and their motion becomes ballistic, following classical trajectories. When a magnetic field *B* is applied perpendicular to the plane, electrons follow cyclotron orbits. Magnetic focusing occurs when electrons injected from one narrow contact focus onto a second contact located an integer number of cyclotron diameters away. By tuning the magnetic field *B* and electron density *n* in the graphene layer, we observe magnetic focusing peaks. We use a cooled scanning gate microscope to image cyclotron trajectories in graphene at 4.2 K. The tip creates a local change in density that casts a shadow by deflecting electrons flowing nearby; an image of flow can be obtained by measuring the transmission between contacts as the tip is raster scanned across the sample. On the first magnetic focusing peak, we image a cyclotron orbit that extends from one contact to the other. In addition, we study the geometry of orbits deflected into the second point contact by the tip.

The unusual properties of graphene offer new approaches to electronics based on the ballistic motion of electrons.[1,2] Covering both sides with hexagonal boron nitride (hBN) sheets greatly enhances the mobility of electrons in graphene devices on a substrate.[3] As a result, electrons can travel several microns without scattering at low temperatures and follow classical trajectories as their motion becomes ballistic.[4,5] Although researchers have observed and studied novel transport phenomena in graphene, such as Klein tunneling,[6,7] specular Andreev reflection,[8,9] Veselago lensing,[10,11] and super-collimation of electron beams in graphene superlattices,[12] direct imaging of trajectories can give us much more information, and the local manipulation of ballistic electrons can open pathways to novel devices. In previous research, we used a cooled scanning probe microscope (SPM) to image electron motion through a two-dimensional electron gas (2DEG) in GaAs/AlGaAs heterostructures.[13-17] We have adapted this technique to image cyclotron orbits in graphene. In a perpendicular magnetic field $B$, electrons travel along cyclotron orbits. Magnetic focusing occurs for electrons travelling from one narrow contact to another, when their spacing $L$ is an integer multiple of the diameter $d_c$ of a cyclotron orbit.[14,16,17] Electrons that enter the graphene sheet at different angles all travel in a circle - as a consequence of this geometry, the electron flux peaks at a distance $d_c$ where circles overlap. As $B$ is increased from zero, the electron transmission from one contact reaches the first magnetic focusing peak when $d_c = L$. As the field continues to increase additional magnetic focusing peaks can occur when $L$ is an integer multiple of $d_c$, if the electron orbit bounces off the edge of the sample specularly.[16,18-20]

In this paper, we present images of the cyclotron orbits in graphene associated with the first magnetic focusing peak, recorded using a cooled SPM at 4.2 K with a tip that acts as a movable gate.[13-17] The sample is a high-mobility hBN-graphene-hBN sandwich patterned into a hall bar geometry using reactive ion etching with a mixture of $CHF_3$ and O (Fig. 1b). The tip, held just above the sample surface, creates an image charge in the graphene that scatters electrons. An image of electron flow is created, by recording the transmission of electrons between the two narrow contacts as the tip is raster scanned across the sample. By tuning the transverse magnetic field $B$ and electron density $n$ in the graphene layer, we observe how the trajectories change as the cyclotron diameter decrease from large values $d_c > L$ at fields below the focusing peak, to smaller values $d_c < L$ above. Unlike conventional materials, the dynamical mass in graphene is density dependent $m^* = \hbar(\pi n)^{1/2}/v_F$ where $v_F$ is the speed associated with the conical bands near the Dirac point;[2,21] the cyclotron diameter is $d_c = 2m^*v_F/eB$. Using SPM imaging, we track the behavior of trajectories on the first magnetic focusing peak as it shifts to higher magnetic fields at higher densities. Previously, magnetic focusing of electrons in a GaAs/AlGaAs 2DEG was imaged using this technique.[16]

**RESULTS AND DISCUSSION**

**Experimental Apparatus.** Figure 1 shows a schematic diagram of the imaging set up and an SEM image of the Hall bar graphene sample. The Hall bar is patterned from a hBN/graphene/hBN sandwich to provide ballistic motion. The Hall bar has dimensions 3.0x4.0 $\mu m^2$, with two narrow (0.7 $\mu$m) contacts along each side, separated by 2.0 $\mu$m and large source and drain contacts at either end. The heavily doped Si substrate acts as a back-gate, covered by a 285 nm insulating layer of $SiO_2$. The back-gate capacitance is $C_G = 11.5$ nF. The density $n$ can be tuned to be either electrons or holes by applying an appropriate voltage $V_G$ between the backgate and the graphene. The density is n = $C_G(V_G - V_{Dirac})/e$ where $e$ is the electron charge and $V_{Dirac}$ is the backgate voltage that nulls the electron density and puts the Fermi level at the Dirac point.

To carry out magnetic focusing measurements, a current source injects a current $I_i$ between contact 1 and the grounded source of the device, as shown in Fig. 1a. Magnetic focusing of electrons between contacts 1 and 2 is sensed, by measuring the voltage $V_s$ between contacts 2 and 3. Because no current can flow into contact 2, the local density and chemical potential change to drive a reverse current into the sample that nulls the total current. The magnetic focusing signal is sensed by $V_s$ and the transresistance $R_m = V_s/I_i$.

**Magnetic Focusing.** Magnetic focusing data are shown in Fig. 2a without the SPM tip present: the transresistance $R_m$ is displayed *vs.* magnetic field $B$ and electron density $n$ at 4.2 K. For conventional semiconductors with parabolic bands, the effective mass $m^*$ is constant and the cyclotron diameter $d_c = 2m^*v/eB$ is determined by the speed $v$ of the carriers. However, for graphene the dynamical mass $m^* = \hbar(\pi n)^{1/2}/v_F$ increases with carrier density $n$, and the speed $v_F$ of electrons is the slope of the conical band, which is fixed near the Dirac point. It follows that the cyclotron diameter $d_c = 2m^*v_F/eB$ increases with density as $n^{1/2}$, and the field for the first magnetic focusing peak is $B_1 = 2m^*/ev_F L$, which increases with density as $n^{1/2}$.

The first magnetic focusing peak is clearly shown in in Fig. 2a which presents experimental measurements of $R_m$ *vs.* $B$ and $n^{1/2}$ at 4.2K with no tip present. As the density and magnetic field are increased, the transresistance peaks (red) along a track with $B_1 \propto n^{1/2}$ as predicted by theory. At magnetic fields $B$ along either side of the magnetic focusing peak, the transmission between the two contacts is reduced (blue), because cyclotron orbit trajectories are focused away from the receiving contact. Evidence for the second magnetic focusing peak with one bounce off the edge between contacts is seen (black) at magnetic fields $B \sim 2B_1$. The intensity of the second peak is reduced by diffuse boundary scattering, which reduces the probability of specular reflection to 0.3 to 0.4 in magnetic focusing measurements[20] in graphene, and to almost zero in 1.0 $\mu$m wide ballistic graphene wires.[22]

**Electron Path Simulations.** Using a simple classical model of electron motion, we simulate electron paths in graphene in a perpendicular magnetic field $B$, including the tip perturbation. The difference in work function between the Si tip and the graphene sample creates an image charge density profile in the graphene sheet:

$$e\Delta n_{tip}(a) = - qh/2\pi(a^2+h^2)^{3/2} \qquad (1)$$

where $a$ is the distance from the tip center, $h$ is the height of the tip above the graphene sheet, and $q$ is the charge on the tip.[13] A peak density change $\Delta n_{tip}(0) = -5 \times 10^{11}$ cm$^{-2}$ at $a = 0$ is chosen to match the data. The density reduction $\Delta n_{tip}(a)$ locally reduces the Fermi energy $E_F(n + \Delta n_{tip})$ while the total chemical potential $E_F(a) + U(a)$ remains constant in space, where $U(a)$ is the potential energy profile created by the tip. In this way, the tip generates a force $F(a) = -\nabla U(a) = \nabla E_F(a)$ on electrons passing nearby that deflects their paths away from the tip position. In graphene, the Fermi energy is $E_F = \hbar v_F(\pi n)^{1/2}$ and the dynamical mass for electron transport is[2,21] $m^* = \hbar(\pi n)^{1/2}/v_F$. This yields the equation of motion for the electron position $r$:

$$d^2r/dt^2 = (1/2)(v_F^2/n)\nabla n(r) \qquad (2)$$

The particle is driven away from areas with lower carrier density beneath the tip.

For each tip position in the simulations, $N = 10,000$ electrons are injected into the sample from one contact at the Fermi energy. The initial position is uniformly distributed across the width of the contact and the initial flux obeys a cosine distribution peaked perpendicular to the contact. The trajectory of each electron through the sample is calculated by digitally integrating Eq. 2, and the transmission $T$ of electrons between contacts 1 and 2 is computed by counting the fraction of emitted trajectories that reach the receiving contact. When present, the tip scatters electron trajectories away from the receiving contact, changing the number received from $p_i$ to $p_{tip}$ and the transmission by $\Delta T = (p_i - p_{tip})/N$. Because electrons can't pass into the receiving contact, the local density and chemical potential build up to create an opposing current that nulls the total flow. In the experiments, the transmission change $\Delta T$ induced by the tip is measured by the voltage change $\Delta V_s$ and the corresponding transresistance change $\Delta R_m = \Delta V_s/I_i$. In these simulations, we assume that trajectories are diffusely scattered by the boundaries, and we neglect possible electrostatic charge accumulation at the sharp edges of the graphene sheet,[23,24] which locally decreases the curvature of cyclotron orbits.

The origin of magnetic focusing is shown by the ray tracing simulations shown in Fig. 2b with no tip present. Electrons leave the injecting contact over a range of angles and circle around cyclotron orbits of diameter $d_c$. The orbits entering at different angles join up a cyclotron diameter $d_c$ away. When the cyclotron diameter $d_c = L$ equals the separation $L$ between contacts, the first magnetic focusing peak in transmission occurs.

The technique to image electron flow with the cooled SPM is illustrated in Fig. 2c, which shows ray tracing simulations at the first focusing peak $B_1$ including the dip in electron density $\Delta n_{tip}(a)$ below the tip, taken from Eq. 1. The dip deflects electrons away from their original orbits, creating a shadow behind the tip location, shown clearly in Fig. 2c. The shadow reduces the electron flow downstream, reducing transmission between contacts when the tip is in a region of strong flow. An image of electron flow is obtained by displaying the transmission change $\Delta T$ from simulations, or the transresistance change $\Delta R_m$ in experiments, as the tip is raster scanned across the sample.

**Experimental Images and Simulations of the Electron Cyclotron Orbit.** Cooled SPM images of electron flow are compared in Fig. 3 for electron density $n = 1.29 \times 10^{12}$ cm$^{-2}$; Fig. 3(a) was recorded at $B = 0$, and Fig. 3(b) was taken at $B_1 = 0.107$ T on the first magnetic focusing peak. In zero magnetic field, no electron flow is visible, but when the field is increased to $B_1$, a clear image of the cyclotron orbit is seen connecting the two contacts. The semicircular arc of negative (red) $\Delta R_m$ results because the tip scatters electrons away from the second contact. When the tip is near the sample edge, the tip can also increase the transmission, as discussed below, resulting in a positive (blue) signal $\Delta R_m$.

For comparison, ray-tracing simulations at the same density $n = 1.29 \times 10^{12}$ cm$^{-2}$ are shown in Fig. 3(c) at $B = 0$ T and in Fig. 3(d) on the first magnetic focusing peak $B_1 = 0.133$ T. The simulations agree well with the SPM images and clearly show a cyclotron orbit connecting the first and second contacts on the first focusing peak. As illustrated by the simulations in Fig. 2c, the tip deflects electron trajectories and creates a shadow downstream that reduces transmission into the receiving contact, generating this image. When the tip is near the sample edge, it can increase transmission by deflecting electrons bound for the edge into the second contact.

We observed how the SPM images of flow and corresponding simulations varied in Fig. 4 by tiling a map of the first magnetic focusing peak in magnetic field $B$ and density $n$ with SPM images of flow between the two contacts in Fig. 4(a) and corresponding simulated images of flow in Fig. 4(b). Cyclotron orbits connecting the two contacts are clearly visible along first focusing peak in $B$ and $n$ shown in Fig. 2(a). To study the effect of increasing the magnetic field at a fixed density, any one row of the tiled plots in Figs. 4(a) and 4(b) can be picked. At field below ($B < B_1$) and above ($B > B_1$) the focusing peak, cyclotron orbits are absent, as one would expect. As $B$ is increased toward the peak, semi-circular cyclotron orbits (red) clearly appear in both the experimental and simulated images. One can track the position of the first magnetic focusing peak in the magnetic field / density map by simply following the location of cyclotron orbit images. Following along the magnetic focusing track in $B_1$ and $n$ we find similar images that show a semi-circular cyclotron orbit of diameter $d_c = L$ from small to high fields and densities, ranging from $B = 0.08$ T and $n = 0.65 \times 10^{12}$ cm$^{-2}$ at the lower left to $B = 0.11$ T and $n = 1.45 \times 10^{12}$ cm$^{-2}$ on the upper right of Fig. 4a, in agreement with a similar sequence shown in the simulations in Fig. 4(b). This consistency supports our interpretation of the image data.

In the experiments and simulations, the region of decreased (red) transmission ($\Delta R_m < 0$ and $\Delta T < 0$) associated with the cyclotron orbit is near the left edge and contacts at lower fields, while a region of enhanced (blue) transmission ($\Delta R_m > 0$ and $\Delta T > 0$) is farther away. As $B$ is increased through the focusing peak at $B_1$, the reduced (red) and enhanced (blue) regions swap places and enhanced transmission now occurs near the edge. One can understand how scattering from the tip acts to increase transmission between the contacts using a simple picture based on the classical cyclotron orbits, using ray-tracing trajectories shown in Fig 5. Figure 5(a) portrays the trajectories when the magnetic field $B = 0.09$T is below the focusing field $B < B_1$ at density $n = 1.29 \times 10^{12}$ cm$^{-2}$: the cyclotron diameter is relatively long ($d_c = 2.9$ $\mu$m $> L$) and the tip can increase transmission inside the sample by tipping orbits back toward the receiving contact. On the contrary, when the magnetic field $B = 0.14$ T at $n = 0.65 \times 10^{12}$ cm$^{-2}$ is above the focusing peak $B > B_1$ in Fig. 5(b), the cyclotron radius is relatively short ($d_c = 1.3$ $\mu$m $< L$), and the tip can increase transmission by tipping orbits away from the diffusely scattering edge, toward the receiving contact. As shown in Fig. 4(a) and 4(b), and the depleted (red) and enhanced (blue) regions swap places as the magnetic field passes through the focusing region and the cyclotron diameter $d_c$ passes through the contact separation $L$.

To study the effect of the electron density $n$ on the cyclotron orbit images, any column of the tiled plots in Figure 4a and 4b can be picked. In both the experimental and simulated images, the magnitude of trans-resistance change $\Delta R_m$ increases at lower $n$, because the tip reduces the electron density below by a fixed amount. At high densities $n > 1\times10^{12}$ cm$^{-2}$, the density dip $\Delta n_{tip}(0) = -5 \times 10^{11}$ cm$^{-2}$ below the tip is relatively small, but at lower densities, the dip becomes comparable to the unperturbed density. This comparison justifies the larger changes in $\Delta R_m$ and $\Delta T$ at lower densities, seen in Figs. 4a and 4b.

**CONCLUSION**

The unique properties of graphene[1,2] open the way for devices based on ballistic electronic transport over distances $\sim 1$ $\mu$m or more. To develop new approaches, we need to learn how electrons travel through ballistic devices. The images above demonstrate how a cooled SPM can image the ballistic flow of electrons through graphene: the capacitively coupled tip deflects electrons, and an image is obtained by displaying the change in transmission between two narrow contacts as the tip is raster scanned across the sample. Similar techniques could be used to track the ballistic motion of electrons through a wide variety of structures.

The cooled SPM could also provide ways to image the motion of electron waves through graphene. Our original imaging experiments on GaAs/GaAlAs 2DEGs showed interference fringes spaced by half the Fermi wavelength, created by the interference of electron waves backscattered by the density depression below the tip[13-15] which allowed us to make an electron interferometer.[15] Although backscattering is reduced in graphene by Klein tunneling,[6,7] electrons can scatter at other angles. A method to observe fringes of electron waves passing through graphene between two point contacts with a cooled SPM was proposed by Braun *et al.*[25] We plan to investigate analogous approaches in the future.

After the experiments were completed, we learned of related work imaging magnetic focusing in graphene.[26]

**METHODS**

**Device Fabrication.** To achieve ballistic transport in graphene (G), we encapsulate it with atomically flat hexagonal-boron nitride (hBN) flakes[27]. The top BN flake is mechanically cleaved onto a polypropylene carbonate (PPC) film spun on a silicon substrate. The PPC with BN flake on is peeled off and transferred onto a Gel Film (Gel-Pak, PF-30/17-X8) sitting on a glass slide. By using a micro-manipulator, the BN/PPC/Gel-Film stamp picks up the graphene, then the bottom BN flake successively, which were then cleaved onto a 285 nm thick silicon oxide ($SiO_2$) substrate. Finally, the BN/G/BN stack is released onto a $SiO_2$ substrate. To achieve highly transparent metallic contacts to the graphene, we expose the freshly etched graphene edge with reactive ion etching and evaporate the chromium and gold electrode immediately afterwards.[11] Here, the electron-beam resist layer serves as an etching mask as well as a lift-off resist layer at the same time.

**Cooled Scanning Probe Microscope.** We use a home-built cooled scanning probe microscope to image the motion of electrons in our sample. The microscope assembly consists of a head assembly where the tip is attached and a cage assembly enclosing the piezotube translator that scans the sample on top in the *X*, *Y* and *Z* directions. Scans are performed by actuating the piezotube with home-built electronics including an *X-Y* position controller for scanning, and a feedback controller for topological scans of the sample surface.[13,16] The microscope assembly is placed in an insert filled with 3.0 mbar of He exchange gas. It is then placed in the bore of a 7 T superconducting solenoid in a liquid helium cryostat that applies a perpendicular magnetic field to the sample. For the transport measurements, standard lock-in amplifiers are used. For the scanning gate measurements, an SPM tip of 20 nm radius was brought at a distance of 10 nm above the BN surface, which is approximately 50 nm above graphene layer. The tip was raster scanned while the trans-resistance $R_m$ was measured.

**AUTHOR INFORMATION**

**Corresponding Author**

Robert M. Westervelt

*Email: westervelt@seas.harvard.edu

**Notes**

The authors declare no competing financial interest.

**ACKNOWLEDGEMENTS**

The authors thank Marko Loncar and Amir Yacoby for helpful discussions. The SPM imaging research and the ray-tracing simulations were supported by the U.S. DOE Office of Basic Energy Sciences, Materials Sciences and Engineering Division, under grant DE-FG02-07ER46422. Graphene sample fabrication was supported by Air Force Office of Scientific Research contract FA9550-13-1-0211. The theory of electron flow was supported by the Science and Technology Center for Integrated Quantum Materials, NSF Grant No. DMR-1231319. Growth of hexagonal boron nitride crystals was supported by the Elemental Strategy Initiative conducted by the MEXT, Japan and a Grant-in-Aid for Scientific Research on Innovative Areas No. 2506 "Science of Atomic Layers" from JSPS. Nanofabrication was performed at the Center for Nanoscale Systems at Harvard, supported in part by an NSF NNIN award ECS-00335765.

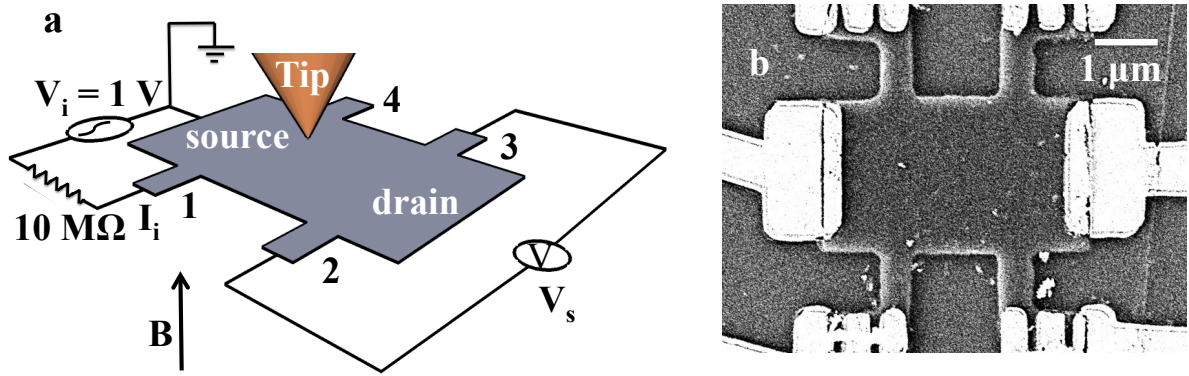

**Figure 1**: (a) Schematic diagram of the experimental setup. An ac current $I_i = 0.1$ uA at 5 kHz is passed between contact 1 and the grounded end source contact, while the voltage difference $V_s$ between contacts 2 and 3 is recorded. The transmission of electrons between contacts 1 and 2 is measured by the transresistance $R_m = V_s/I_i$. The tip of a cooled scanning probe microscope capacitively creates a dip in the electron density below. To image the electron flow between contacts 1 and 2, the signal $R_m$ is displayed while the tip is raster scanned across the sample at a constant height (20 nm). (b) Scanning electron micrograph of the Hall bar sample constructed from a hBN/graphene/hBN sandwich.

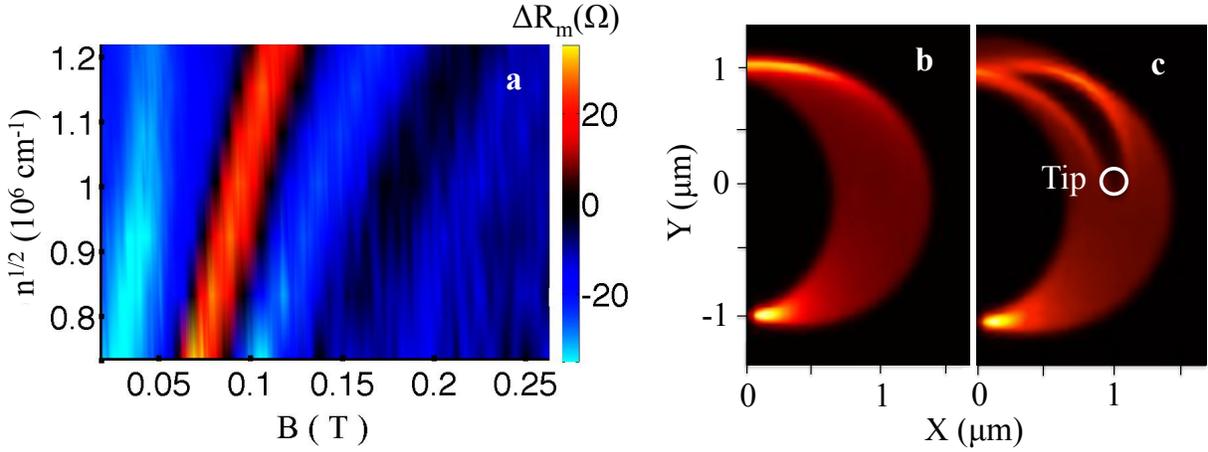

**Figure 2**: (a) Measured transresistance $\Delta R_m$ displayed *vs.* magnetic field $B$ and square root of electron density $n^{1/2}$ at 4.2 K. The first magnetic focusing peak where cyclotron orbits connect contacts 1 and 2 is clearly shown as the region of enhanced $R_m$ (red) bordered by regions of reduced $R_m$ (blue) where the orbits miss the second contact. Signs of a second magnetic focusing peak at twice the magnetic field are also shown. (b) Ray-tracing simulation shows the origins of the first magnetic focusing peak when the cyclotron orbits connect the two contacts located at $X = 0$ and $Y = \pm 1$ $\mu$m. (c) Ray-tracing simulation shows scattering by the density depression immediately below the SPM tip that creates a shadow on the second point contact. For (b) and (c) $B = 0.133$ T and $n = 1.29 \times 10^{12}$ cm$^{-2}$.

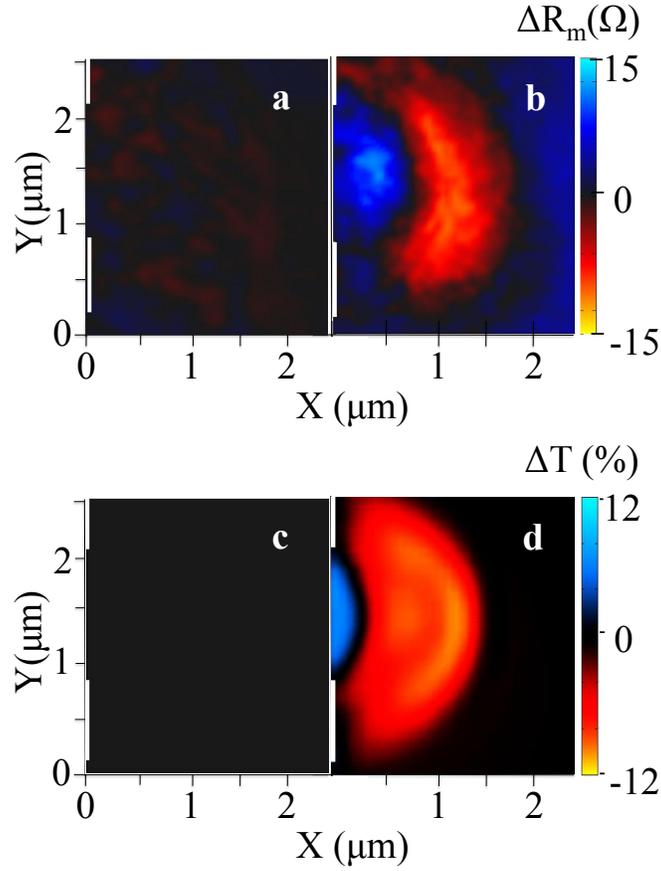

**Figure 3**: SPM images of electron flow through graphene at 4.2 K for density $n = 1.29 \times 10^{12}$ cm$^{-2}$. The sample edge is at $X = 0$ and the contacts are shown as bold white lines. (a) SPM image in zero magnetic field $B = 0$; no electron flow is seen. b) SPM image of flow on the first magnetic focusing peak that shows the cyclotron orbit joining the two contacts as semicircular paths of reduced transresistance $\Delta R_m < 0$. A region of enhanced $\Delta R_m > 0$ is also seen near the sample edge (see below). (c) Ray-tracing simulations of transmission between contacts for $B = 0$; no flow is seen. (d) Ray-tracing transmission change $\Delta T$ on the first magnetic focusing peak corresponding to panel (b) shows the semi-circular cyclotron orbit ($\Delta T < 0$) as well as a region enhanced flow ($\Delta T > 0$) near the sample edge.

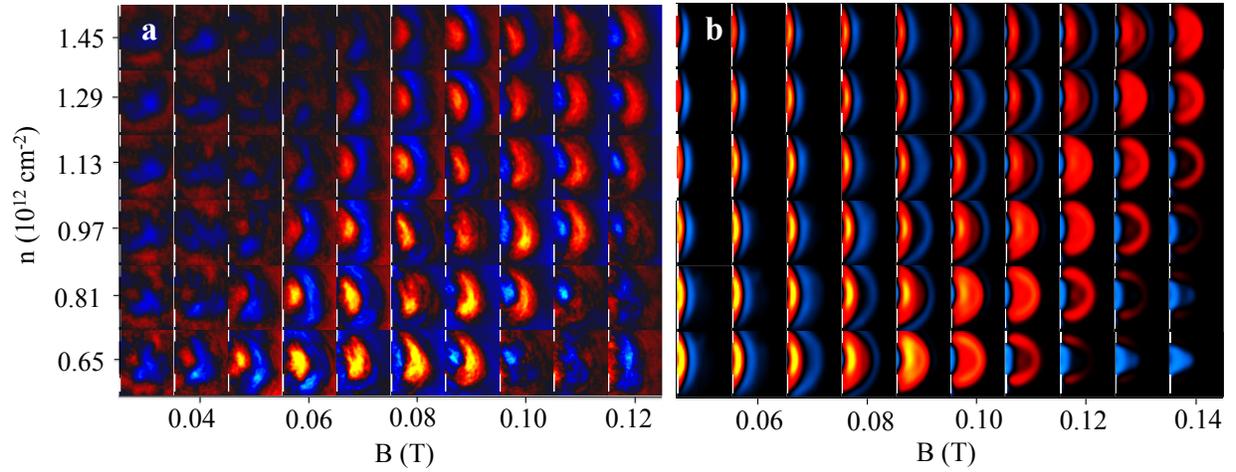

**Figure 4**: Tiled images of (a) SPM images of electron flow ($\Delta R_m$) between contacts 1 and 2 and (b) ray-tracing simulations of flow ($\Delta T$) as the magnetic field $B$ and the electron density $n$ are varied over the first magnetic focusing peak $B_1(n)$; the $B$ range for simulations is shifted slightly to cover the focusing peak. For $B$ and $n$ near the magnetic focusing peak, cyclotron orbits with diameters $d_c = L$ that connect the two contacts are clearly shown in the (a) experiments and (b) simulations. For $B < B_1$ a region of enhanced transmission (blue) appears toward the right, away from the contacts, as the tip knocks long cyclotron orbits back into the receiving contact. As the field increases to values $B > B_1$ the blue region switches toward the left edge, as the tip bounces electrons with short cyclotron orbits away from the wall and into the receiving contact. The $X$ and $Y$ axes and the color maps are given in Fig. 3b for (a) the SPM image panels and in Fig. 3(d) for (b) the simulation panels.

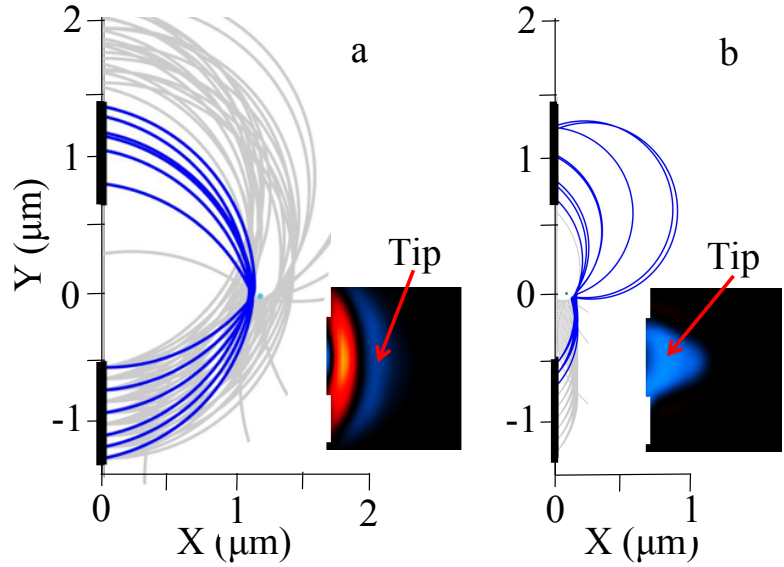

Figure 5: Simulated electron trajectories for regions of enhanced transmission at low and high magnetic fields: (a) Ray-tracing trajectories at $B = 0.09$ T and $n = 1.29 \times 10^{12}$ cm$^{-2}$ below the first magnetic focusing peak. The cyclotron diameter 2.86 um is larger than the contact spacing, and most trajectories miss the receiving contact. However, the transmission increases when the tip deflects rays toward the receiving contact (blue rays). (b) Electron trajectories at $B = 0.14$ T and $n = 0.65 \times 10^{12}$ cm$^{-2}$ above the first magnetic focusing peak. The cyclotron diameter 1.3 um is smaller than the contact spacing, and these trajectories are diffusely scattered by the edge and do not contribute to the transmission. The transmission increases when the tip deflects rays away from the edge and toward the receiving contact (blue rays).